\documentstyle[11pt,epsf]{article}
\begin{document}
\title{
Low-Temperature Series for the Square Lattice Potts Model 
by the Improved Finite-Lattice Method
       }
\author{
        H. ARISUE \\ 
        Osaka Prefectural College of Technology \\ 
        Saiwai-cho, Neyagawa, Osaka 572, Japan
              \and
        K. TABATA \\
        Osaka Institute of Technology \\ 
        Junior College  \\ 
        Ohmiya, Asahi-ku, Osaka 535, Japan
        }
\maketitle

\begin{abstract}
The low-temperature series are calculated 
for the free energy, magnetization and susceptibility 
in the $Q$-state Potts model on the square lattice,
using the improved algorithm of the finite lattice method.
The series are obtained to the order of $z^{41}$ for each of $Q=5-50$,
and the result of their Pad\'e type analysis is compared 
with those of the large-$Q$ expansion and the Monte Carlo simulations.
\end{abstract}

\newpage
\section{Introduction}

 The finite-lattice method 
is a very efficient technique to generate 
the expansion series in statistical models \cite{Enting1977,Creutz}
and in lattice gauge theory \cite{Arisue1984}.
 In this method the expansion series for the free energy density 
in the infinite-volume limit is given 
by an appropriate linear combination of 
the free energies on finite-size lattices. 
The partition function of the finite-size lattices, 
from which the free energy of the lattices is obtained,
can be calculated using the transfer matrix formalism \cite{Enting1980}.
 This method avoids the problem involved in the graphical method, 
in which it is rather difficult 
to give the algorithm for listing all the diagrams 
that contribute to the relevant order of the series. 
  The maximum order of the expansion series is determined 
by the maximum size of the lattices 
whose partition function can be calculated exactly to the order. 

 In the case of the model whose spin-variable at each site takes 
many discrete values, 
however, one can only calculate the partition function of 
the finite-size lattices whose sizes are relatively small
and it is difficult to obtain the series to such high order 
as in the Ising model, in which the spin variable takes only two values.
 Recently one of the authors \cite{Arisue1995b} proposed 
an improved algorithm of the finite-lattice method, 
which enables one to obtain higher order terms 
than the standard finite-lattice method 
in the spin systems whose spin variables take 
a large or infinite number of discrete values. 
 It was applied to the low-temperature expansion 
for the free energy of the solid-on-solid (SOS) model and 
it enabled him to extend the series to the order of $u^{23}$ 
\cite{Arisue1995b}
from the order of $u^{12}$ that was
obtained using the standard finite-lattice method
\cite{Hasenbusch1994}.

 In this paper we apply this improved finite lattice method 
to the calculation of the low-temperature series
for the free energy, magnetization and susceptibility 
in the $Q$-state Potts model on the square lattice.
This system has the first order phase transition for $Q \ge 5$
\cite{Potts}.
The values of the free energy, latent heat, and 
the gap of the specific heat and magnetization at the critical point
are known exactly \cite{Baxter},
while the exact values of the specific heat itself and the magnetic 
susceptibility at the critical point are not known yet. 
We calculate the series to order $z^{41}$ for each of $ Q=5-50$. 
We can extend the series from the ones calculated previously 
by Bhanot {\it et al} \cite{Bhanot1993} and Briggs {\it et al}
\cite{Briggs}
using the standard finite-lattice method,
to substantially higher orders for large values of $Q \leq 10$,
and our series for $11 \leq  Q \leq 50$ 
have the order that is 
about 3 times higher than the ones given by Kihara {\it et al}
for general $Q$ and zero field
\cite{Kihara}
or by Straley and Fisher
\cite{Straley}
for general $Q$ and general field.
We analyze the obtained series 
using the inhomogeneous differential approximants
and their integration.
It serves to check the validity of the series obtained
and of the method of the analysis used
for the quantities whose values are known exactly.
On the other hand, 
it makes a new prediction for the quantities 
whose exact values are not known.

In section 2, 
we describe the improved algorithm of the finite-lattice method 
for the low-temperature expansion 
of the $Q$-state Potts model on the square lattice. 
The series obtained by the improved algorithm 
of the finite-lattice method are given in section 3.
We present the analysis of the series in section 4.
The result of our analysis is compared with those of the large-$Q$ 
expansion and the Monte Carlo simulations.
Our results are summarized in section 5.

\section{Algorithm}
 Here we give the improved algorithm of the finite-lattice method 
for the low-temperature expansion 
of the $Q$-state Potts model on the square lattice, 
which enables us to obtain longer series than 
the standard finite-lattice method.
 Let us consider the two-dimensional 
$L_x \times L_y$ rectangular lattice $\Lambda_0$. 
The free energy density in the infinite-volume limit is given by 
\begin{equation}
     f =  
   - \lim_{L_x,L_y \rightarrow \infty} 
          \frac{1}{ L_x L_y } 
         \ln{ \big[ Z(\Lambda_0;Q) \big] },
\end{equation}
with 
\begin{equation}
    Z(\Lambda_0;Q)
       =  \sum_{\{0 \le s_i \le Q-1\}} 
          \exp{ \big[ - \beta \sum_{<i,j>} (1-\delta_{s_i,s_j})  
                      - h \sum_{i} ( 1 - \delta_{s_i,0}) \big] }
\end{equation}
where the integer variable $s_i$ at each site $i$ of $\Lambda_0$ is 
restricted to $ 0 \le s_i \le Q-1$. 
 The low-temperature series is calculated 
with respect to the expansion parameter $z=\exp{(-\beta)}$. 
 We take the boundary condition 
that all the variables outside $\Lambda_0$ are fixed to be zero. 

 We consider the set $\{\Lambda\}$ 
of all two-dimensional rectangular sub-lattices of $\Lambda_0$. 
 The sub-lattice $\Lambda$ is denoted by its size $l_x \times l_y $ 
and its position in $\Lambda_0$.
 We define a set of integers $\xi_q$ that consist of 
$0$ and $(q-1)$ integers among the set of integers $\{1,2,\dots,Q-1\}$. 
 We define the $H$ of $\Lambda$ and $\xi_q$ as 
\begin{equation}
   H(\Lambda;\xi_q) 
     = - \ln{ \big[ Z(\Lambda;\xi_q) \ \big] }.     
      \label{defH}
\end{equation}
In the calculation of the partition function 
$Z(\Lambda;\xi_q)$,
the variable $s_i$ at each site $i$ inside $\Lambda$ is restricted 
to be one of the elements of $\xi_q$,
and all the variables outside $\Lambda$ are fixed to be zero. 
We define $W$ of $\Lambda$ and $\xi_q$ recursively as
\begin{equation}
    W(\Lambda;\xi_q)
       = H(\Lambda;\xi_q) 
       -   \sum_{\scriptstyle
                    \Lambda^{'} \subseteq \Lambda, 
               \atop\scriptstyle 
                   \xi_{q^{'}} \subseteq \xi_q
                  }^{ \hspace{0.8cm} '} \hspace{-0.0cm}
           W( \Lambda^{'} ; \xi_{q^{'}} ). 
        \label{defW}
\end{equation}
 Here the prime in the summation ${\sum}^{'}$ implies that 
the $W(\Lambda'=\Lambda;\xi_{q^{'}}=\xi_{q})$ should be excluded
in taking the summation.
 We note that the $H(\Lambda;\xi_q)$ and $W(\Lambda;\xi_q)$ 
depend on the size of $\Lambda$ but not on its position 
and depend on the value of $q$
but not on the details of the elements of the $\xi_q$.
 So we denote them as $H(l_x \times l_y;q)$ and $W(l_x \times l_y;q)$,
then we can rewrite Eq.~(\ref{defW}) as
\begin{eqnarray}
&&    W(l_x \times l_y;q)
       = H(l_x \times l_y;q) \nonumber\\
&&
        - \sum_{\scriptstyle
                 {
                    l_x^{'} \le l_x, l_y^{'} \le l_y, 
               \atop\scriptstyle 
                 2 \le q^{'} \le q }
                  }^{ \hspace{0.8cm} '} \hspace{-0.0cm}
          (l_x-l_x^{'}+1)(l_y-l_y^{'}+1)
                  \left( \begin{array}{c} q-1 \\
                                  q'-1 \end{array}  \right)
           W( l_x^{'} \times l_y^{'} ; q^{'} ). 
           \label{eqW}
\end{eqnarray}
 We know 
\begin{eqnarray}
   H(L_x \times L_y;Q) 
    &=& \sum_{\scriptstyle
                l_x \le L_x, \ l_y \le L_y, 
                    \atop\scriptstyle 
                         \ q \le Q }  
                         (L_x-l_x+1)(L_y-l_y+1)    
            \left( \begin{array}{c} Q-1 \\  q-1 \end{array}  \right)
                                                    \nonumber \\
    && \;\;\;\;\;\;\;\;\;\;\;\;\;\;\;\;\;\;\;\;\;\;\;\;\;\;\;\;\;
       \;\;\;\;\;\;\;\;\;\;\;\;\;
                \times W( l_x \times l_y;q ).  
\end{eqnarray}
 Taking the infinite-volume limit we obtain 
\begin{eqnarray}
    f &=&  \lim_{ L_x,L_y \rightarrow \infty} 
            \frac{1}{ L_x L_y} H(L_x \times L_y;Q)    \nonumber \\
      &=&  \sum_{ l_x,l_y,  q \le Q} 
            \left( \begin{array}{c} Q-1 \\  q-1 \end{array}  \right)
                  W( l_x \times l_y;q ).                
      \label{free_energy}
\end{eqnarray}
 The magnetization $M$ and susceptibility $\chi$ are given 
from this free energy density, 
by 
\begin{equation}
    M = 1 - \frac{Q}{Q-1} \frac{\partial f}{\partial h} 
\end{equation}
and 
\begin{equation}
    \chi = \frac{\partial^2 f}{\partial h^2}.
\end{equation}
 In the practical calculation we introduce
the variable $x \equiv \exp{(h)}-1$ \cite{Guttmann1993}, then
\begin{equation}
     \frac{\partial}{\partial h} \mbox{\LARGE $\vert$}_{h=0}
           = \frac{\partial}{\partial x} \mbox{\LARGE $\vert$}_{x=0}
\end{equation}
and 
\begin{equation}
     \frac{\partial^2}{\partial h^2} \mbox{\LARGE $\vert$}_{h=0}
           = (\frac{\partial^2}{\partial x^2}
             + \frac{\partial}{\partial x})\mbox{\LARGE $\vert$}_{h=0}
\end{equation}
and it is enough to evaluate the expansion series 
for the free energy density to the second order
in $x$ in order to obtain the zero-field magnetization and 
susceptibility.

 In the standard cluster expansion of the free energy 
\cite{Domb1974,Muenster1981} for this model, 
a cluster is composed of polymers and  
each of the polymers consists of the excited sites 
that are connected by nearest-neighbor bonds. 
 An integer value $s_i \in \{ 1, 2, \dots,  Q-1 \}$ is put to each 
excited site $i$ of the polymer. 
 For each cluster we can define
the subset of the set $\{ 1, 2, \dots,  Q-1 \}$ so that 
each element of this subset is found to be put on 
at least one of the excited sites of all the polymers 
the cluster is composed of.
Then we can assign to the cluster an integer $\tilde{q}$
so that $(\tilde{q}-1)$ is the number of the elements of the subset.
 We can prove 
\cite{Arisue1984} 
that the Taylor expansion of the $W(l_x \times l_y;q)$ 
with respect to $z$ includes 
the contribution from all the clusters of polymers 
in the standard cluster expansion that have $\tilde{q} = q$ 
and that can be embedded into the $l_x \times l_y$ lattice 
but cannot be embedded into any of its rectangular sub-lattices.

 So the series expansion of the $W(l_x \times l_y;q)$ 
starts from the order $z^{n(l_x,l_y,q)}$ with 
$n(l_x,l_y,q) $ given in the following. 
It is enough to consider only for $l_x \le l_y$.
We first define $k \equiv [(q-1)/(l_x l_y)]$,
where we denote $[p]$ as the maximum integer that is less than $p$.
In the case of $k=0$ ( namely $q-1 \le l_x l_y$), 
the cluster that contributes to the lowest-order term
of the expansion series of the $W(l_x \times l_y;q)$
consists of a single polymer. 
We know in this case that

\ \hfill \raisebox{-.65ex}{\( \displaystyle
n(l_x,l_y,q)= { \Biggl\{ }
\)}
\hspace{-1.6in}
\parbox{\textwidth}{
\begin{eqnarray}
  \ \ \ \ \ \ \ \ \ \ 
           & 2 (l_x+l_y)+q-2 & \mbox{ for \ \ \ $q \le l_x+l_y  $ }  
                    \label{eqA} \\
  \ \ \ \ \ \ \ \ \ \ 
           & l_x+l_y+2 q-2   & \mbox{ for \ \ \ $q > l_x+l_y $ }.
                    \label{eqB}
\end{eqnarray}
}

\noindent
Examples of the clusters that correspond to these two cases 
in Eqs.~(\ref{eqA}) and (\ref{eqB}) are presented 
in Fig.1 (a) and (b), respectively. 
In the figures the integers denote the excited spin variables 
at the sites and the solid lines represent the excited bonds.
In the case of $k \ge 1$ ( namely $q-1 > l_x l_y $ ),  
the cluster that contributes to the lowest order term
is composed of $k+1$ polymers, 
among which a polymer consists of $q' \equiv q-1-k l_x l_y$ sites
and each of the other $k$ polymers consists of the $l_x \times l_y$ sites. 
Then if $q' \le {l_x}^2$ 

\ \hfill \raisebox{.50ex}{\( \displaystyle
n(l_x,l_y,q)= { \Biggl\{ }
\)}
\hspace{-1.6in}
\parbox{\textwidth}{
\begin{eqnarray}
  \ \ \ \ \ \ \ \ \ \ \ \ \ \ \ \ \ \ 
           & k \{l_x (l_y+1)+(l_x+1) l_y\} +2 q'+2 i+1 \nonumber \\ 
           & \mbox{ \ \ \ \ \ \ \ \ \ \ \ \ \ 
                \          for \ \ \ $q' \le i(i+1)$ }  \label{eqC} \\
  \ \ \ \ \ \ \ \ \ \ \ \ \ \ \ \ \ \
           & k \{l_x (l_y+1)+(l_x+1) l_y\} +2 q'+2 i+2 \nonumber \\
           & \mbox{ \ \ \ \ \ \ \ \ \ \ \ \ \ 
                         for \ \ \ $q' > i(i+1)$ } \label{eqD} 
\end{eqnarray}
}

\noindent
where  $i \equiv [\sqrt{q'}\ ]$ .
Examples of the clusters that correspond to these two cases 
in Eqs.~(\ref{eqC}) and (\ref{eqD}) are presented 
in Fig.2 (a) and (b), respectively. 
If $q' > {l_x}^2$ 
\begin{equation}
       n(l_x,l_y,q)=k \{l_x (l_y+1)+(l_x+1) l_y\}
                         + 2 q'+2 l_x+j+1
                 \label{eqE}
\end{equation}
where $j \equiv [(q'-l_x^2)/l_x]$.
Example of the cluster that corresponds to Eq.~(\ref{eqE}) 
is presented in Fig.2 (c).
 To obtain the expansion series to order $z^N$,
we should take into account all the combinations of $l_x$, $l_y$ 
and $q$
that satisfy 
$  n(l_x,l_y,q) \le N $
in the summation of Eq.~(\ref{free_energy}) 
and should evaluate each of the $W$'s to the order of $z^N$. 

 We should stress that the improved algorithm described above 
enables us to calculate the series for large $Q$ that is substantially longer 
than the case when one would apply the standard finite-lattice method 
naively \cite{Bhanot1993,Briggs}. 
 In the latter case one would define 
\begin{equation}
   \tilde{H}(\Lambda;\xi_Q) 
     = - \ln{ \big[ Z(\Lambda;\xi_Q) \ \big] }     
      \label{defHprime}
\end{equation}
and 
\begin{equation}
  \tilde{W}(\Lambda;\xi_Q)
    = \tilde{H}(\Lambda;\xi_Q) 
     - {\sum_{\scriptstyle \Lambda ' \subset \Lambda } }
      \tilde{W}( \Lambda ';\xi_Q) 
\end{equation}
for the rectangular $l_x \times l_y$ sub-lattice $\Lambda$. 
Here the variable $s_i$ takes all integer values 
$ 0 \le s_i \le Q-1 $ for each site $i \in \Lambda$ 
in the calculation 
of the partition function $Z(\Lambda;\xi_Q)$. 
 Then the free energy density in the infinite-volume limit 
would be given by 
\begin{equation}
    f =  \sum_{ l_x,l_y} \tilde{W}( l_x \times l_y ;\xi_Q).
      \label{free_energy2}      
\end{equation}
 The series expansion of this $\tilde{W}(l_x \times l_y;\xi_Q)$ 
starts from the order of $z^{2(l_x + l_y)}$.
The clusters that give the contribution to this order 
are the clusters consisting of a single polymer,
an example of which is depicted in Fig.3.
 So for the series expansion to order $z^N$, 
one should take into account all the rectangular lattices 
that satisfy 
$ 2(l_x + l_y) \le N $
in the summation of Eq.~(\ref{free_energy2}). 
The improved algorithm might appear to be more complicated 
than this standard algorithm, 
since the former involves the summation with respect to $q$ 
as well as the summation with respect to $l_x \times l_y$. 
The CPU time and memory for evaluating the partition functions 
of the finite-size lattices needed to obtain the series 
to the same order are, however, much smaller in the improved algorithm. 
In order to obtain the series to $z^{41}$, for instance, 
we should calculate the partition functions 
up to the lattice-size of $10 \times 10$
either in the standard algorithm or in the improved algorithm. 
In the standard algorithm, 
the partition function with $q=Q$ 
is needed for this maximum size of the lattice.
On the other hand, in the improved algorithm 
the partition function with $q=2$ is enough 
for this size of the lattice. 

\section{Series}
 Using the improved algorithm of the finite-lattice method 
described in the previous section 
we have calculated the low-temperature series for the free energy
density $f$, magnetization $M$ and susceptibility $\chi$ 
in the zero field for the Potts model on the square lattice 
to order $z^{41}$ for $Q=5-50$. 
 The obtained coefficients are listed 
in Table~\ref{tab:coeff20} and \ref{tab:coeff50} only for $Q=20$ and $50$,
where we give the series for $z=\exp{(f)}$ 
instead of the free energy density $f$, 
as well as the series for the magnetization and susceptibility. 
If the reader would like to know the coefficients for the other values of $Q$,
he can get them from the authors by e-mail or 
from the authors' web-site.\cite{Address}  
The calculation was performed by FACOM-VPP500 at KEK (Tsukuba)
and FACOM-VP2600 at Kyoto University Data Processing Center.
 The maximum size of the used main memory was about 500Mbytes.
 We have checked that each of the $W(l_x \times l_y;q)$'s 
in eq.~(\ref{free_energy}) starts from the correct order of
$z^{n(l_x,l_y,q)}$ described in the previous section.

 The series were obtained 
previously using the standard finite-lattice method 
by Bhanot {\it et al} \cite{Bhanot1993} 
to order $z^{25}$ for $Q=8$ and  
by Briggs {\it et al} \cite{Briggs} to order
$z^{39}$, $z^{39}$, $z^{39}$, $z^{35}$, $z^{31}$ and  $z^{31}$ 
for $Q=5,6,7,8,9$ and $10$, respectively.
 The series were also calculated by Kihara {\it et al}\cite{Kihara} 
to order $z^{16}$ for general $Q$  and zero field
and by Straley and Fisher\cite{Straley} 
for general $Q$ and general field to order $z^{13}$
using the graphical method. 
 The series coefficients obtained here by us are consistent 
with the coefficients obtained by these people to their order. 

 We note that in the standard finite-lattice method 
the maximum order of the obtainable expansion series
reduces when one increases the value of the $Q$,
while the improved algorithm enables us to calculate the series 
to the same order for all the $Q$ 
that are larger than or equal to some value.
 In fact, we can calculate the series to the order of $u^{41}$ 
for an arbitrarily large $Q ( > 17)$ 
using the $W(l_x,l_y;q)$'s that have been evaluated here 
to obtain the series to order $u^{41}$ for $Q=17$,
although we have stopped it at $Q=50$.
 This can be understood by noticing 
from Eqs.~(\ref{eqA})-(\ref{eqE}) that 
the series expansion of the $W(l_x,l_y;q)$ 
for arbitrary $l_x$, $l_y$ and $q$ with $q > 17$
starts from the order that is higher than $z^{41}$ 
and does not contribute to the expansion series 
to the order of $z^{41}$.

\def\ltsim{\matrix{<\cr\noalign{\vskip-7pt}\sim\cr}}
\def\gtsim{\matrix{>\cr\noalign{\vskip-7pt}\sim\cr}}
\section{Analysis of the series}

The $Q$-state Potts model with $Q > 4$ has 
a first-order phase transition.
For the first-order phase transition, some important quantities,
such as the susceptibility and specific heat, have finite values
at the critical point, in contrast with the fact that these values
are infinite for the second-order phase transition.

The infiniteness of the critical value 
for the second-order phase transition
sometimes makes the analysis easier; if the physical
quantity would have a strong singularity, we would be able to determine 
the critical exponent with a satisfactory accuracy 
by the differential approximations such as the Pad\'e analysis.

On the other hand, in the case of the first-order phase transition,
we have to integrate the differential approximants
to determine the critical amplitude.
This integration, however, has a subtle problem discussed below,
and as a result, we must choose the approximants carefully.

In analyzing the expansion series, we use homogeneous or inhomogeneous
first order differential approximants \cite{IDA}, in which the
approximants to a function $f(z)$ satisfy
\begin{equation}
    Q_1(z) f'(z) + Q_2(z) f(z) = R(z).  \label{inhomo}
\end{equation}
Here, $Q_1(z)$, $Q_2(z)$ and $R(z)$ are polynomials
of order $M_1$, $M_2$, and $M$, respectively,
which are determined so that Eq.~(\ref{inhomo}) is satisfied 
to the order of $z^{N}$ with $N$ the maximum order 
of the expansion series of the $f(z)$.
If $R(z)$ is identically zero, the approximant is just the
D-log Pad\'e approximant.

To obtain the value of $f(z)$ at the critical point $z=z_c$,
we must integrate
the function $f'(z) = \{ - Q_2(z) f(z) + R(z) \} / Q_1(z)$
from 0 to $z_c$.
For the first order phase transition,
the denominator $Q_1(z)$ of $f'(z)$  has in most cases 
a zero close to the critical point. 
The zero point is usually slightly above the critical point 
on the real axis of $z$. 
It is sometimes, however, below the critical point, 
then the integral of $f'(z)$ will diverge and we should abandon such 
an approximant.
Even if the zero point is above the critical point, 
it is so close to the critical point that 
its subtle fluctuation makes 
the fluctuation of the critical value $f(z_c)$ rather large.


At first, in order to examine the accuracy of the analysis 
of the series described above, 
we apply it to the free energy density $f_c$ 
at the critical point ($z_c = 1/(1+\sqrt{Q})$),
the latent heat $\Delta U$ and the magnetization gap $\Delta M$,
whose exact values were obtained by Baxter \cite{Baxter}.
In Table~\ref{tab:estimates} the estimates of the free energy density,
latent heat and magnetization gap from the longest series ($N=41$)
are presented
for $Q \geq 7$, and
their exact values are in parentheses.
These estimates obtained by the analysis of the series
may contain errors by two reasons; One of them is 
the statistical fluctuation among the approximants
with the different orders $M_1, M_2$ and $M$ of the polynomials 
$Q_1(z)$ and $Q_2(z)$ in Eq.~(\ref{inhomo}) 
but with the same order $N$ of the original series.
About these statistical errors we find two common tendency;
i) If the order $M$ is too large, the convergence of the approximants 
is not so good, and 
if the orders $M_1$ and $M_2$ are too small, 
the integral of $f'(z)$ diverges frequently,
therefore we restrict the order $M$ to be $M=-1$ (D-log Pad\'e approximant) 
or $M=0$ -- 6 (inhomogeneous differential approximants) and the orders
$M_1$ and $M_2$ to satisfy $|M_1 - M_2| \le 10$.
ii) For smaller $Q$, some approximants are finite at the critical point
but apparently far from the exact value, then we must exclude them.
We present only the statistical errors in Table~\ref{tab:estimates}.
The other error is from the finiteness of the series.
If there is discrepancy between an estimate for the finite series
and the corresponding exact value beyond the statistical error 
for each $N$, 
we may say that the discrepancy is caused by
the finiteness of the series and the error arising from the finiteness
of the series has the order of magnitude of this discrepancy.

Now, we present the result of the individual analysis 
for the free energy density,
the latent heat and the magnetization gap.

For the free energy density $f_c$, 
both the inhomogeneous and homogeneous
differential approximants give well-converged results, 
the latter converging more excellently.
The data in Table~\ref{tab:estimates} are from the latter.
For $Q \gtsim 30$, the estimates from the D-log Pad\'e approximants
agree with the exact value within one standard deviation.
On the other hand, for $Q \ltsim 20$, the estimates are 
larger than the exact value, and do not agree with it
within one standard deviation.
As an example in Fig.4, we present a plot of the estimates for $Q=20$ versus
the number $N$ of terms of the truncated series.
The difference between the estimates and the exact value is
about 0.01 percent while the fluctuation of the approximants 
is about 0.005 percent, the former is about 2 times larger than the latter.
The ratio of the difference between the estimate and the exact value 
to the statistical fluctuation becomes larger for smaller $Q$.
Nevertheless, the figure also exhibits the tendency that the difference 
between the estimates and
the exact value decreases with the increase of the number of terms.
(We should mention that the errorbars for $35 \le N \le 38$ are 
relatively small. It is because, for these $N$, 
some approximants have divergent integral of $f'(z)$
and the remaining fewer approximants happen to be very close.)
Therefore, this difference can be expected
to shrink for longer series.
It would, however,  be difficult to extrapolate the data 
to $N \rightarrow \infty$ to estimate the exact value, 
if it were not known, since we do not know
what asymptotic behavior the approximants should follow with respect to $N$.

For the latent heat, 
the inhomogeneous differential approximants
give more convergent results
than the D-log Pad\'e approximants, 
in contrast with the case of the free energy density.
Then, by using the inhomogeneous differential approximants,
we can obtain about 60 estimates of the latent heat
for each model with $Q \geq 7$.
For $Q \gtsim 40$, the estimates coincide with the exact value
within one standard deviation.
For $Q \ltsim 30$, although the estimates do not agree 
with the exact value
within one standard deviation, 
we find that their difference becomes smaller for larger number of terms 
of the expansion series and so
we can say that the difference would disappear for long enough series,
as in the previous case of the free energy density.

For the magnetization gap,
about 10 estimates obtained by the D-log Pad\'e approximants 
converge well for each $Q$,
while those obtained by the inhomogeneous differential approximants
do not.
Although it does not agree with the exact value within one standard
deviation except for $Q \simeq 50$, 
the estimates themselves are very close to the exact value;
for example, in the case of $Q=20$,
the difference between the estimates and the exact value is
about 0.015 percent.
Further Fig.5 shows that the differences
would disappear for larger number of terms as well as in the case of
the free energy density and the latent heat.

Now using the same method as the above,
we estimate the values of the susceptibility and
specific heat at the critical point,
whose exact values are not known.

The behavior of the differential approximants for the susceptibility
is like that for the magnetization gap.
The D-log Pad\'e approximants for the susceptibility show better
convergence than the inhomogeneous approximants,
although the errors of the estimates for the susceptibility are
larger than those for the magnetization gap.
The estimates from the longest series ($N=41$) are 
listed in Table~\ref{tab:estimates2},
and our estimates for $7 \leq Q \leq 10$ are consistent with 
the estimates obtained by Briggs {\it et al}\cite{Briggs}.
For $Q=10$, 15, 20,
Monte Carlo estimates are given by Janke {\it et al}\cite{Janke95}.
As their definition of the susceptibility is different from ours
in the normalization factor,
the results in Ref~\cite{Janke95} should multiplied by a factor
of $\{q/(q-1)\}^2$
to fit into our definition.
The estimates obtained from Ref~\cite{Janke95} are
$3.777 \pm 0.035$ ($Q=10$), $0.7052 \pm 0.0033$ ($Q=15$)
and $0.30242 \pm 0.00050$ ($Q=20$).
Our corresponding estimates are $2.78 \pm 0.14$ ($Q=10$),
$0.657 \pm 0.017$ ($Q=15$) and $0.2938 \pm 0.0036$ ($Q=20$).
Although there are differences over one standard deviation,
which are especially large for $Q=10$, we may expect that
the differences would disappear for larger $Q$.

For the specific heat,
the estimates obtained by the D-log Pad\'e approximants give well
converging results,
while the inhomogeneous differential approximants
do not.
We list the estimates from the longest series ($N=41$) by the D-log Pad\'e
approximants in Table~\ref{tab:estimates3}.
The errors of the estimates for the specific heat are
smaller than those for the susceptibility.
Nevertheless, 
our estimates for $7 \leq Q \leq 10$ are inconsistent with 
the estimates obtained by Briggs {\it et al}\cite{Briggs}. 
Therefore we should check 
the validity of our approximants in some way.

Fortunately, by the duality relation \cite{Baxter}, 
we can obtain the exact value of 
the gap of the specific heat as follows\cite{Baxter};
\begin{equation}
C_{v}^{+} - C_{v}^{-} = \frac{\beta ^2 \Delta U}{\sqrt{Q}} . \label{eq:gap}
\end{equation}
In Table~\ref{tab:estimates3}, we also list the estimates of
the gap of the specific heat, where the specific heat $C_{v}^{+}$ 
is calculated using the high-temperature expansion series 
obtained from the low-temperature series by the duality.
It is clear that for $Q \geq 11$
there exists the gap of the specific heat
and more over almost all the estimated values 
of the gap of the specific heat coincide with
the exact values given by Eq.~(\ref{eq:gap}) within the standard deviation.
Therefore, we convince the validity of our estimates 
for the specific heat.

Finally we compare our estimates of the specific heat at the critical
point with the estimates obtained from the large-$Q$ expansion series
that was recently obtained by Bhattacharya {\it et al} \cite{Bhattacharya},
and with the ones obtained from Monte Carlo simulations
\cite{Billoire92,Billoire93,Janke92,Rummukainen,Janke95}.
In Table~\ref{tab:estimates4}, we show our estimates of the specific
heat again and the estimates obtained from the large-$Q$ expansion
series and the Monte Carlo simulations.
For $Q = 7,\ 8,\ 10,\ 15,\ 20,\ 30$, 
the values of the estimates from the large-$Q$ expansion are
sited in Ref~\cite{Bhattacharya}, the others are
calculated by the regularized logarithmic Pad\'e approximants
given in Ref~\cite{Bhattacharya}.
Among the Monte Carlo results those of Ref.~\cite{Janke95}
have small statistical errors and they are consistent with 
the results of the large-$Q$ expansion for $Q=10,15,20$.
Our estimates from the low-temperature series
are systematically smaller than the estimates from the large-$Q$ expansion.
In Fig.6 and 7, we present a plot of the estimates 
of the specific heat $C_v^{-}$ for $Q=20$ and $Q=50$, respectively, 
versus the number $N$ of the terms of the series truncated.
We can see that the estimates are approaching the result of the large-$Q$ 
expansion from below for each $Q$ and that they approach faster 
for larger $Q$.
Thus we can expect that, if we extrapolate the estimates in some way 
to $N \rightarrow \infty$ for each $Q$, 
we would obtain the estimates that may be more consistent with the 
result of the large-$Q$ expansion, 
although we do not know how to make the extrapolation correctly, 
as mentioned above.
%

\section{Summary}
We obtained the low-temperature series 
for the free energy, magnetization and susceptibility
of the Potts model on the square lattice.

Using an improved algorithm of the finite lattice method,
we extended the series from those given by the
standard algorithm of the finite lattice method.
Our improved algorithm is more efficient 
for the higher-state Potts models.

Using the new series, we calculated the critical values 
of the free energy density, latent heat, magnetization gap, 
susceptibility and specific heat.

The estimated values of the former three are very close to the
known exact values.
The values of the latter two, 
which are not known exactly, are obtained for
various $Q$ with high precision.
Especially, the estimates of the specific heat are very close
to the estimates from the large $Q$ expansion series.


\scriptsize
\begin{table}[htb]
\caption{
The low-temperature expansion coefficients
for the free energy, magnetization and
susceptibility for the $Q= 20$ 
square lattice Potts model.
         }
\label{tab:coeff20}
\begin{center}
\begin{tabular}{|r|r|r|r|}
\hline
    $n$  & \multicolumn{1}{c|}{$z$} 
         & \multicolumn{1}{c|}{$M$}
         & \multicolumn{1}{c|}{$\chi$}  \\
\hline
$  0$ & $                         1$ & $                           1$ & $                           0$ \\ 
$  1$ & $                         0$ & $                           0$ & $                           0$ \\ 
$  2$ & $                         0$ & $                           0$ & $                           0$ \\ 
$  3$ & $                         0$ & $                           0$ & $                           0$ \\ 
$  4$ & $                        19$ & $                         -20$ & $                          19$ \\ 
$  5$ & $                         0$ & $                           0$ & $                           0$ \\ 
$  6$ & $                        38$ & $                         -80$ & $                         152$ \\ 
$  7$ & $                       684$ & $                       -1440$ & $                        2736$ \\ 
$  8$ & $                      -589$ & $                        1460$ & $                       -2280$ \\ 
$  9$ & $                      4104$ & $                      -12960$ & $                       36936$ \\ 
$ 10$ & $                     34466$ & $                     -109520$ & $                      323912$ \\ 
$ 11$ & $                    -46512$ & $                      138240$ & $                     -199728$ \\ 
$ 12$ & $                    483398$ & $                    -2040420$ & $                     7678261$ \\ 
$ 13$ & $                   1614240$ & $                    -7086240$ & $                    30536496$ \\ 
$ 14$ & $                  -2010808$ & $                     4671200$ & $                    19059128$ \\ 
$ 15$ & $                  48638556$ & $                  -259282080$ & $                  1242232920$ \\ 
$ 16$ & $                  66750002$ & $                  -417704060$ & $                  2800285550$ \\ 
$ 17$ & $                  94851648$ & $                 -1155653280$ & $                 11160075600$ \\ 
$ 18$ & $                4264583402$ & $                -27837269920$ & $                166147691080$ \\ 
$ 19$ & $                2711283660$ & $                -28967068800$ & $                314891078616$ \\ 
$ 20$ & $               36660260676$ & $               -326575493500$ & $               2700423490565$ \\ 
$ 21$ & $              350530337736$ & $              -2773117509120$ & $              20580457476816$ \\ 
$ 22$ & $              226902725452$ & $              -3504413044080$ & $              47276982916744$ \\ 
$ 23$ & $             5653101156048$ & $             -54127038428640$ & $             481746412796832$ \\ 
$ 24$ & $            28469510466547$ & $            -273739383621540$ & $            2536124771313858$ \\ 
$ 25$ & $            42704598074544$ & $            -626534197990560$ & $            8388189200399472$ \\ 
$ 26$ & $           700159265957050$ & $           -7500005076811440$ & $           75083308577493088$ \\ 
$ 27$ & $          2423368661949468$ & $          -28449325073826720$ & $          325937254907403792$ \\ 
$ 28$ & $          7731033540174246$ & $         -110572911429835580$ & $         1480698106281593115$ \\ 
$ 29$ & $         78953345947533348$ & $         -956758345892651520$ & $        10938227296679669016$ \\ 
$ 30$ & $        233171802367220678$ & $        -3290695663943877920$ & $        45003875885372649120$ \\ 
$ 31$ & $       1207925854600942620$ & $       -17857155839057961120$ & $       248795624942616383064$ \\ 
$ 32$ & $       8578891752427054593$ & $      -118198071379878114580$ & $      1550417616117914081584$ \\ 
$ 33$ & $      26279052771743353332$ & $      -427018427440815394080$ & $      6633731834136735494256$ \\ 
$ 34$ & $     169448419303276320478$ & $     -2669246990084147723840$ & $     39743309408406296543112$ \\ 
$ 35$ & $     934014456459287376408$ & $    -14635750107536528768640$ & $    219466703191297256269272$ \\ 
$ 36$ & $    3342530604057094050901$ & $    -59864493412976212314860$ & $   1016877625830611545531997$ \\ 
$ 37$ & $   22227466452797269128528$ & $   -378821839108831184988000$ & $   6119224442354906298876816$ \\ 
$ 38$ & $  105216422923458425776276$ & $  -1862780050724742025689040$ & $  31557668972616720906032400$ \\ 
$ 39$ & $  449173545929305443593316$ & $  -8659232479851255144790560$ & $ 157936029967724328819066816$ \\ 
$ 40$ & $ 2809270008061904464069897$ & $ -52184954811154800744150940$ & $ 921011351332383770376427924$ \\ 
$ 41$ & $12477877823064242776774212$ & $-246220110417995423090111520$ & $4633614244759932587057525712$ \\ 
\hline
\end{tabular}
\end{center}
\end{table}
\clearpage

\tiny
\begin{table}[htb]
\caption{
The low-temperature expansion coefficients
for the free energy, magnetization and
susceptibility for the $Q= 50$ 
square lattice Potts model.
         }
\label{tab:coeff50}
\begin{center}
\begin{tabular}{|r|r|r|r|}
\hline
    $n$  & \multicolumn{1}{c|}{$z$} 
         & \multicolumn{1}{c|}{$M$}
         & \multicolumn{1}{c|}{$\chi$}  \\
\hline
$  0$ & $                              1$ & $                                 1$ & $                                 0$ \\ 
$  1$ & $                              0$ & $                                 0$ & $                                 0$ \\ 
$  2$ & $                              0$ & $                                 0$ & $                                 0$ \\ 
$  3$ & $                              0$ & $                                 0$ & $                                 0$ \\ 
$  4$ & $                             49$ & $                               -50$ & $                                49$ \\ 
$  5$ & $                              0$ & $                                 0$ & $                                 0$ \\ 
$  6$ & $                             98$ & $                              -200$ & $                               392$ \\ 
$  7$ & $                           4704$ & $                             -9600$ & $                             18816$ \\ 
$  8$ & $                          -4459$ & $                             11150$ & $                            -20580$ \\ 
$  9$ & $                          28224$ & $                            -86400$ & $                            254016$ \\ 
$ 10$ & $                         659246$ & $                          -2019800$ & $                           6003872$ \\ 
$ 11$ & $                       -1025472$ & $                           3225600$ & $                          -7018368$ \\ 
$ 12$ & $                       12281948$ & $                         -50077050$ & $                         193485271$ \\ 
$ 13$ & $                       89987520$ & $                        -370953600$ & $                        1510567296$ \\ 
$ 14$ & $                     -162228808$ & $                         567974000$ & $                        -886343752$ \\ 
$ 15$ & $                     3720812256$ & $                      -19045411200$ & $                       92683335360$ \\ 
$ 16$ & $                    12381176822$ & $                      -67093148150$ & $                      376836950000$ \\ 
$ 17$ & $                     4469044608$ & $                      -89631139200$ & $                     1066326898560$ \\ 
$ 18$ & $                   895089855002$ & $                    -5548225826800$ & $                    32926830187720$ \\ 
$ 19$ & $                  1892364280800$ & $                   -13708568736000$ & $                   107149526869056$ \\ 
$ 20$ & $                 13423422398316$ & $                  -111911089584250$ & $                   924768638863685$ \\ 
$ 21$ & $                206844895493376$ & $                 -1536698755564800$ & $                 11106335245654656$ \\ 
$ 22$ & $                478049132085952$ & $                 -4409226660382200$ & $                 42663913301510704$ \\ 
$ 23$ & $               5667428458933248$ & $                -51068413852233600$ & $                452593396732200192$ \\ 
$ 24$ & $              49346646734282527$ & $               -435539546923202850$ & $               3808046989807658388$ \\ 
$ 25$ & $             172491504405475584$ & $              -1825893233439542400$ & $              19671214965561745152$ \\ 
$ 26$ & $            1938967535168345710$ & $             -19591446789689634600$ & $             194639282941507768168$ \\ 
$ 27$ & $           13089424692665334048$ & $            -135834659906560636800$ & $            1410212431991986624512$ \\ 
$ 28$ & $           63921872954358377616$ & $            -750734284293452018450$ & $            8836405252541575381155$ \\ 
$ 29$ & $          628985731889425518048$ & $           -7147786492756957612800$ & $           80157906305935393997376$ \\ 
$ 30$ & $         3839485035589391850158$ & $          -45982543354085211648800$ & $          551763520397297550084960$ \\ 
$ 31$ & $        23463259463706802592160$ & $         -302788074318283787164800$ & $         3888163728504881482313664$ \\ 
$ 32$ & $       203694326578125205781553$ & $        -2595462149212992934427950$ & $        32749085439406082880398944$ \\ 
$ 33$ & $      1225877434229949054953952$ & $       -16606999340819591568259200$ & $       224791274766846749241616896$ \\ 
$ 34$ & $      8555010161977020587142598$ & $      -120783311584253799401189600$ & $      1691607730655033745533673792$ \\ 
$ 35$ & $     66800901314888122476718848$ & $      -948877081581250695129561600$ & $     13374510959732980701033577152$ \\ 
$ 36$ & $    417354805969242736779031351$ & $     -6276444566058217793336109650$ & $     93992501752971816579399237857$ \\ 
$ 37$ & $   3081300147534452125687872768$ & $    -47482109079734479233388368000$ & $    725062658320307588405549126016$ \\ 
$ 38$ & $  22520630415621757023630981256$ & $   -353898756249751908192492010600$ & $   5521526378626835675327811477000$ \\ 
$ 39$ & $ 148545447485043078418570354656$ & $  -2445977768521719097911967382400$ & $  39997172375744231349328962685056$ \\ 
$ 40$ & $1106109615352046378739786168307$ & $ -18559451854582591017244861914850$ & $ 308483818208942470837183549866604$ \\ 
$ 41$ & $7848119276197068664163827052832$ & $-135219203614185545984625196732800$ & $2311827629935984338141853691434752$ \\ 
\hline
\end{tabular}
\end{center}
\end{table}
\clearpage
\normalsize
\begin{table}[hbt]
\caption{Estimates of the free energy density $f_c$,
the latent heat $\Delta U$ and
the magnetization gap $\Delta M\;$.}
\label{tab:estimates}
\begin{center}
\begin{tabular}{cccc} \hline
$Q$ &        $f_c$             & $\Delta U$          & $\Delta M$    \\ \hline
 7 & $-0.0430048 \pm 0.0000024$ & $0.35720 \pm 0.00883$ & $0.7681 \pm 0.0060$\\
   &($-0.0431119$)              & (0.35328)             & (0.749565)        \\
 8 & $-0.0392927 \pm 0.0000055$ & $0.49058 \pm 0.00193$ & $0.8080 \pm 0.0056$\\
   &($-0.0393533$)              & (0.48636)             & (0.799837)         \\
 9 & $-0.0360255 \pm 0.0000044$ & $0.60375 \pm 0.00219$ & $0.8394 \pm 0.0020$\\
   &($-0.0360805$)              & (0.59967)             & (0.833261)        \\
10 & $-0.0332047 \pm 0.0000035$ & $0.69832 \pm 0.00182$ & $0.8605 \pm 0.0020$\\
   &($-0.0332365$)              & (0.69605)             & (0.857107)        \\
11 & $-0.0307291 \pm 0.0000024$ & $0.78047 \pm 0.00057$ & $0.87746 \pm 0.00072$ 
\\
   &($-0.0307586$)              & (0.77860)             & (0.875107)        \\
12 & $-0.0285722 \pm 0.0000025$ & $0.85122 \pm 0.00027$ & $0.89053 \pm 0.00047$ 
\\
   &($-0.0285900$)              & (0.84994)             & (0.888878)        \\
13 & $-0.0266652 \pm 0.0000023$ & $0.91316 \pm 0.00029$ & $0.90114 \pm 0.00032$ 
\\
   &($-0.0266817$)              & (0.91216)             & (0.899999)         \\
14 & $-0.0249831 \pm 0.0000021$ & $0.96756 \pm 0.00012$ & $0.90989 \pm 0.00025$ 
\\
   &($-0.0249931$)              & (0.96689)             & (0.909086)         \\
15 & $-0.0234812 \pm 0.0000018$ & $1.015966 \pm 0.000090$ & $0.92094 \pm 
0.00124$ \\
   &($-0.0234908$)              & (1.015414)             & (0.916664)      \\
16 & $-0.0221416 \pm 0.0000018$ & $1.059128 \pm 0.000067$ & $0.92351 \pm 
0.00018$ \\
   &($-0.0221471$)              & (1.058749)             & (0.923075)      \\
17 & $-0.0209338 \pm 0.0000015$ & $1.098023 \pm 0.000054$ & $0.92888 \pm 
0.000012$ \\
   &($-0.0209393$)              & (1.097700)             & (0.928570)      \\
18 & $-0.0198439 \pm 0.0000013$ & $1.133140 \pm 0.000041$ & $0.93358 \pm 
0.00012$ \\
   &($-0.0198486$)              & (1.132916)             & (0.933332)       \\
19 & $-0.0188560 \pm 0.0000011$ & $1.165129 \pm 0.000038$ & $0.937683 \pm 
0.000094$ \\
   &($-0.0188594$)              & (1.164923)             & (0.937499)       \\
20 & $-0.0179567 \pm 0.0000010$ & $1.194298 \pm 0.000036$ & $0.941318 \pm 
0.000008$ \\
   &($-0.0179586$)              & (1.194155)             & (0.941176)       \\
30 & $-0.0120422 \pm 0.0000004$ & $1.390466 \pm 0.000011$ & $0.962987 \pm 
0.000006$ \\
   &($-0.0120425$)              & (1.390442)             & (0.962963)       \\
40 & $-0.0089727 \pm 0.0000002$ & $1.498232 \pm 0.000008$ & $0.972980 \pm 
0.000004$ \\
   &($-0.0089727$)              & (1.498224)             & (0.972930)       \\
50 & $-0.0071096 \pm 0.0000001$ & $1.567525 \pm 0.000003$ & $0.978726 \pm 
0.000001$ \\
   &($-0.0071096$)              & (1.567522)             & (0.978723)       \\ \hline
\end{tabular}
\end{center}
\end{table}

\begin{table}[hbt]
\caption{Estimates of the susceptibility $\chi_c\;$.}
\label{tab:estimates2}
\begin{center}
\begin{tabular}{ccc} \hline
$Q$ & $\chi(z_c)$            \\ \hline
 5 & 154      $\pm$ 19       \\ 
 6 &  49.5    $\pm$  4.1     \\ 
 7 &  19.0    $\pm$  1.3     \\ 
 8 &   8.09   $\pm$  0.29    \\ 
 9 &   4.55   $\pm$  0.20    \\ 
10 &   2.78   $\pm$  0.14    \\ 
11 &   1.900  $\pm$  0.047   \\ 
12 &   1.372  $\pm$  0.026   \\ 
13 &   1.036  $\pm$  0.027   \\ 
14 &   0.815  $\pm$  0.019   \\ 
15 &   0.657  $\pm$  0.017   \\ 
16 &   0.539  $\pm$  0.016   \\ 
17 &   0.4490 $\pm$  0.0080  \\ 
18 &   0.3844 $\pm$  0.0061  \\ 
19 &   0.3347 $\pm$  0.0047  \\ 
20 &   0.2938 $\pm$  0.0036  \\
30 &   0.1195 $\pm$  0.0002  \\
40 &   0.0697 $\pm$  0.0002  \\
50 &   0.0478 $\pm$  0.00004 \\ \hline
\end{tabular}
\end{center}
\end{table}
\clearpage
\begin{table}[hbt]
\caption{Estimates of the specific heat.}
\label{tab:estimates3}
\begin{center}
\begin{tabular}{ccccc} \hline
$Q$ & $C_v ^-$ & $C_v ^+$ & $\Delta C_v$ (series) & $\Delta C_v$ (exact) \\%
\hline
 7 & 52.98   $\pm$ 0.61   & 68.01   $\pm$ 0.47   & 15.0    $\pm$ 0.76   & 
0.2234 \\
 8 & 28.47   $\pm$ 0.47   & 30.17   $\pm$ 0.06   &  1.70   $\pm$ 0.48   & 
0.3099 \\
 9 & 20.03   $\pm$ 0.27   & 20.87   $\pm$ 0.05   &  0.84   $\pm$ 0.27   & 
0.3841 \\
10 & 15.579   $\pm$ 0.083   & 16.294   $\pm$ 0.034   &  0.715   $\pm$ 0.090   & 
0.4476 \\
11 & 12.856   $\pm$ 0.061   & 13.493   $\pm$ 0.034   &  0.637   $\pm$ 0.070   & 
0.5021 \\
12 & 10.985   $\pm$ 0.050   & 11.622   $\pm$ 0.055   &  0.636   $\pm$ 0.074   & 
0.5492 \\
13 &  9.598   $\pm$ 0.032   & 10.258   $\pm$ 0.062   &  0.659   $\pm$ 0.070   & 
0.5901 \\
14 &  8.548   $\pm$ 0.027   &  9.204   $\pm$ 0.020   &  0.656   $\pm$ 0.034   & 
0.6260 \\
15 &  7.721  $\pm$ 0.028  &  8.388  $\pm$ 0.004  &  0.666  $\pm$ 0.029  & 
0.6576 \\
16 &  7.044  $\pm$ 0.011  &  7.739  $\pm$ 0.002  &  0.695  $\pm$ 0.011  & 
0.6856 \\
17 &  6.4908  $\pm$ 0.0086  &  7.2061  $\pm$ 0.0041  &  0.715  $\pm$ 0.010  & 
0.7106 \\
18 &  6.0251  $\pm$ 0.0074  &  6.7637  $\pm$ 0.0095  &  0.739  $\pm$ 0.012  & 
0.7330 \\
19 &  5.6317  $\pm$ 0.0061  &  6.3859  $\pm$ 0.0098  &  0.754  $\pm$ 0.012  & 
0.7532 \\
20 &  5.2913  $\pm$ 0.0053  &  6.0645  $\pm$ 0.0069  &  0.773  $\pm$ 0.009  & 
0.7714 \\
30 &  3.4012 $\pm$ 0.0010 &  4.2919 $\pm$ 0.0051 &  0.8906 $\pm$ 0.0052 & 
0.8861 \\
40 &  2.5864 $\pm$ 0.0006 &  3.5303 $\pm$ 0.0019 &  0.9439 $\pm$ 0.0020 & 
0.9393 \\
50 &  2.1218 $\pm$ 0.0004 &  3.0936 $\pm$ 0.0007 &  0.9717 $\pm$ 0.0008 & 
0.9668 \\ \hline
\end{tabular}
\end{center}
\end{table}
\clearpage
\begin{table}[hbt]
\caption{Estimates of the specific heat: low-temperature
series vs. large $Q$ series and Monte Carlo simulations.}
\label{tab:estimates4}
\begin{center}
\begin{tabular}{cccc} \hline

$Q$ & low-temperature series & large $Q$ series & Monte Carlo simulations \\ \hline
 7 & 52.98   $\pm$ 0.61   & 69.6      $\pm$ 0.5      & 50 $\pm$ 10 \cite{Billoire92} \\
   &                      &                          & 47.5 $\pm$ 2.4 \cite{Janke92} \\
   &                      &                          & 44.4 $\pm$ 2.2 \cite{Rummukainen} \\
 8 & 28.47   $\pm$ 0.47   & 36.9      $\pm$ 0.2      & \\
 9 & 20.03   $\pm$ 0.27   & 24.1      $\pm$ 0.2      & \\
10 & 15.579  $\pm$ 0.083  & 17.98     $\pm$ 0.02     & 12.7 $\pm$ 0.3 \cite{Billoire92} \\
   &                      &                          & 17.81 $\pm$ 0.10 \cite{Janke95} \\
11 & 12.856  $\pm$ 0.061  & 14.24     $\pm$ 0.05     & \\
12 & 10.985  $\pm$ 0.050  & 11.84     $\pm$ 0.02     & \\
13 &  9.598  $\pm$ 0.032  & 10.17     $\pm$ 0.01     & \\
14 &  8.548  $\pm$ 0.027  &  8.939    $\pm$ 0.009    & \\
15 &  7.721  $\pm$ 0.028  &  7.999    $\pm$ 0.003    & 8.004 $\pm$ 0.019 \cite{Janke95} \\
16 &  7.044  $\pm$ 0.011  &  7.248    $\pm$ 0.004    & \\
17 &  6.4908 $\pm$ 0.0086 &  6.643    $\pm$ 0.003    & \\
18 &  6.0251 $\pm$ 0.0074 &  6.142    $\pm$ 0.002    & \\
19 &  5.6317 $\pm$ 0.0061 &  5.720    $\pm$ 0.001    & \\
20 &  5.2913 $\pm$ 0.0053 &  5.3612   $\pm$ 0.0004   & 5.2 $\pm$ 0.2 \cite{Billoire92} \\
   &                      &                          & 5.3612 $\pm$ 0.0055 \cite{Janke95} \\
30 &  3.4012 $\pm$ 0.0010 &  3.41294  $\pm$ 0.00005  & \\
40 &  2.5864 $\pm$ 0.0006 &  2.58986  $\pm$ 0.00002  & \\
50 &  2.1218 $\pm$ 0.0004 &  2.123198 $\pm$ 0.000006 & \\
\hline

\end{tabular}
\end{center}
\end{table}
\clearpage

\centerline{\bf Figure Captions}

\vspace{5mm}
Fig. 1: Examples of the cluster consisting of a single polymer ( $k=0$) 
that contributes to the lowest-order term of the expansion series
for (a) $W(l_x=4,l_y=5;q=4)$ and (b) $W(l_x=4,l_y=5;q=11)$. 

\vspace{5mm}
Fig. 2: Examples of the cluster consisting of two polymers ( $k=1$) 
that contributes to the lowest-order term of the expansion series 
for (a) $W(l_x=3,l_y=4;q=18)$, (b) $W(l_x=3,l_y=4;q=21)$ 
and (c)$W(l_x=3,l_y=4;q=24)$. 

\vspace{5mm}
Fig. 3: Examples of the cluster consisting of a single polymer 
that contributes to the lowest-order term of the expansion series
for $\tilde{W}(l_x=4,l_y=5;\xi_Q)$.

\vspace{5mm}
Fig. 4: The estimates of the free energy density $f_c$ for $Q=20$ versus
the number of terms of the series. The solid line shows the exact
value.

\vspace{5mm}
Fig. 5:  The estimates of the magnetization gap $\Delta M$ for $Q=20$ versus
the number of terms of the series. The solid line shows the exact
value.

\vspace{5mm}
Fig. 6: The estimates of the specific heat $C_v^-$ for $Q=20$ versus
the number of terms of the series. The solid line shows the estimates
obtained from the large-$Q$ expansion series by using the regularized
logarithmic Pad\'e approximants.

\vspace{5mm}
Fig. 7: The estimates of the specific heat $C_v^-$ for $Q=50$ versus
the number of terms of the series. The solid line shows the estimates
obtained from the large-$Q$ expansion series by using the regularized
logarithmic Pad\'e approximants.
\clearpage
\newpage
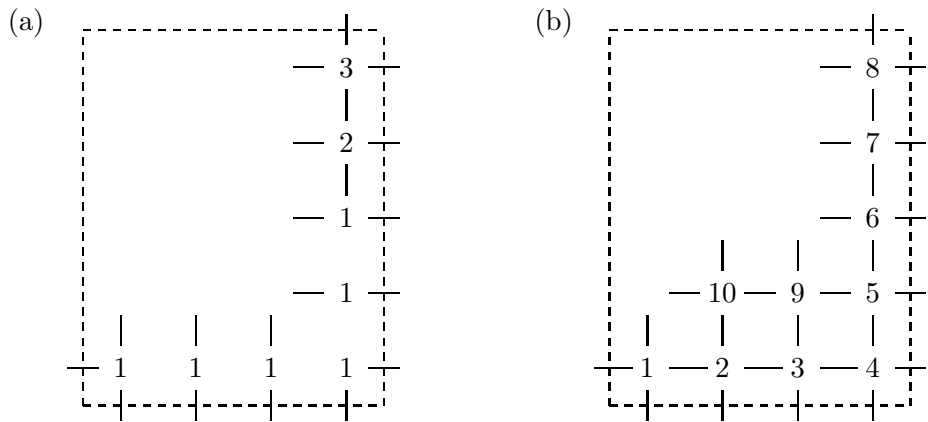
\begin{figure}
\setlength{\unitlength}{1mm}
\begin{picture}(110,60)
\put(0,0){\begin{picture}(50,60)(-5,0)
\put(-5,55){{(a)}}
\multiput(5,5)(2,0){20}{\line(1,0){1}}
\multiput(5,5)(0,2){25}{\line(0,1){1}}
\multiput(5,55)(2,0){20}{\line(1,0){1}}
\multiput(45,5)(0,2){25}{\line(0,1){1}}
\multiput(10,10)(10,0){4}{\makebox(0,0){1}}
\multiput(40,20)(0,10){2}{\makebox(0,0){1}}
\put(40,40){\makebox(0,0){2}}
\put(40,50){\makebox(0,0){3}}
\put(3,10){\line(1,0){4}}
\multiput(10,3)(10,0){4}{\line(0,1){4}}
\multiput(10,13)(10,0){3}{\line(0,1){4}}
\multiput(33,20)(0,10){4}{\line(1,0){4}}
\multiput(43,10)(0,10){5}{\line(1,0){4}}
\multiput(40,33)(0,10){3}{\line(0,1){4}}
\end{picture}
}
\put(70,0){\begin{picture}(50,60)(-5,0)
\put(-5,55){{(b)}}
\multiput(5,5)(2,0){20}{\line(1,0){1}}
\multiput(5,5)(0,2){25}{\line(0,1){1}}
\multiput(5,55)(2,0){20}{\line(1,0){1}}
\multiput(45,5)(0,2){25}{\line(0,1){1}}
\put(10,10){\makebox(0,0){1}}
\put(20,10){\makebox(0,0){2}}
\put(30,10){\makebox(0,0){3}}
\put(40,10){\makebox(0,0){4}}
\put(40,20){\makebox(0,0){5}}
\put(40,30){\makebox(0,0){6}}
\put(40,40){\makebox(0,0){7}}
\put(40,50){\makebox(0,0){8}}
\put(30,20){\makebox(0,0){9}}
\put(20,20){\makebox(0,0){10}}
\multiput(3,10)(10,0){4}{\line(1,0){5}}
\multiput(10,3)(10,0){4}{\line(0,1){4}}
\multiput(10,13)(10,0){4}{\line(0,1){4}}
\multiput(33,10)(0,10){5}{\line(1,0){4}}
\multiput(43,10)(0,10){5}{\line(1,0){4}}
\multiput(40,3)(0,10){6}{\line(0,1){4}}
\multiput(13,20)(10,0){2}{\line(1,0){4}}
\multiput(20,23)(10,0){2}{\line(0,1){4}}
\end{picture}
}
\end{picture}
\caption{
Examples of the cluster consisting of a single polymer ( $k=0$) 
that contributes to the lowest-order term of the expansion series
for (a) $W(l_x=4,l_y=5;q=4)$ and (b) $W(l_x=4,l_y=5;q=11)$. }
\end{figure}
\begin{figure}
\setlength{\unitlength}{1mm}
\begin{picture}(100,160)
\put(0,110){\begin{picture}(100,50)(-5,0)
\put(-5,45){{(a)}}
\multiput(5,5)(2,0){15}{\line(1,0){1}}
\multiput(5,5)(0,2){20}{\line(0,1){1}}
\multiput(5,45)(2,0){15}{\line(1,0){1}}
\multiput(35,5)(0,2){20}{\line(0,1){1}}
\put(10,10){\makebox(0,0){1}}
\put(20,10){\makebox(0,0){2}}
\put(30,10){\makebox(0,0){3}}
\put(10,20){\makebox(0,0){4}}
\put(20,20){\makebox(0,0){5}}
\put(30,20){\makebox(0,0){6}}
\put(10,30){\makebox(0,0){7}}
\put(20,30){\makebox(0,0){8}}
\put(30,30){\makebox(0,0){9}}
\put(10,40){\makebox(0,0){10}}
\put(20,40){\makebox(0,0){11}}
\put(30,40){\makebox(0,0){12}}
\multiput(3,10)(10,0){4}{\line(1,0){4}}
\multiput(3,20)(10,0){4}{\line(1,0){4}}
\multiput(3,30)(10,0){4}{\line(1,0){4}}
\multiput(3,40)(10,0){4}{\line(1,0){4}}

\multiput(10,3)(10,0){3}{\line(0,1){4}}
\multiput(10,13)(10,0){3}{\line(0,1){4}}
\multiput(10,23)(10,0){3}{\line(0,1){4}}
\multiput(10,33)(10,0){3}{\line(0,1){4}}
\multiput(10,43)(10,0){3}{\line(0,1){4}}
\put(45,25){\makebox(0,0){+}}
\multiput(55,5)(2,0){15}{\line(1,0){1}}
\multiput(55,5)(0,2){20}{\line(0,1){1}}
\multiput(55,45)(2,0){15}{\line(1,0){1}}
\multiput(85,5)(0,2){20}{\line(0,1){1}}
\put(60,10){\makebox(0,0){13}}
\put(70,10){\makebox(0,0){14}}
\put(70,20){\makebox(0,0){15}}
\put(60,20){\makebox(0,0){16}}
\put(80,10){\makebox(0,0){17}}
\multiput(53,10)(10,0){4}{\line(1,0){4}}
\multiput(53,20)(10,0){3}{\line(1,0){4}}

\multiput(60,3)(10,0){3}{\line(0,1){4}}
\multiput(60,13)(10,0){3}{\line(0,1){4}}
\multiput(60,23)(10,0){2}{\line(0,1){4}}
\end{picture}
}
\put(0,55){\begin{picture}(100,50)(-5,0)
\put(-5,45){{(b)}}
\multiput(5,5)(2,0){15}{\line(1,0){1}}
\multiput(5,5)(0,2){20}{\line(0,1){1}}
\multiput(5,45)(2,0){15}{\line(1,0){1}}
\multiput(35,5)(0,2){20}{\line(0,1){1}}
\put(10,10){\makebox(0,0){1}}
\put(20,10){\makebox(0,0){2}}
\put(30,10){\makebox(0,0){3}}
\put(10,20){\makebox(0,0){4}}
\put(20,20){\makebox(0,0){5}}
\put(30,20){\makebox(0,0){6}}
\put(10,30){\makebox(0,0){7}}
\put(20,30){\makebox(0,0){8}}
\put(30,30){\makebox(0,0){9}}
\put(10,40){\makebox(0,0){10}}
\put(20,40){\makebox(0,0){11}}
\put(30,40){\makebox(0,0){12}}
\multiput(3,10)(10,0){4}{\line(1,0){4}}
\multiput(3,20)(10,0){4}{\line(1,0){4}}
\multiput(3,30)(10,0){4}{\line(1,0){4}}
\multiput(3,40)(10,0){4}{\line(1,0){4}}

\multiput(10,3)(10,0){3}{\line(0,1){4}}
\multiput(10,13)(10,0){3}{\line(0,1){4}}
\multiput(10,23)(10,0){3}{\line(0,1){4}}
\multiput(10,33)(10,0){3}{\line(0,1){4}}
\multiput(10,43)(10,0){3}{\line(0,1){4}}
\put(45,25){\makebox(0,0){+}}
\multiput(55,5)(2,0){15}{\line(1,0){1}}
\multiput(55,5)(0,2){20}{\line(0,1){1}}
\multiput(55,45)(2,0){15}{\line(1,0){1}}
\multiput(85,5)(0,2){20}{\line(0,1){1}}
\put(60,10){\makebox(0,0){13}}
\put(70,10){\makebox(0,0){14}}
\put(70,20){\makebox(0,0){15}}
\put(60,20){\makebox(0,0){16}}
\put(80,10){\makebox(0,0){17}}
\put(80,20){\makebox(0,0){18}}
\put(80,30){\makebox(0,0){19}}
\put(70,30){\makebox(0,0){20}}
\multiput(53,10)(10,0){4}{\line(1,0){4}}
\multiput(53,20)(10,0){4}{\line(1,0){4}}
\multiput(63,30)(10,0){3}{\line(1,0){4}}

\multiput(60,3)(10,0){3}{\line(0,1){4}}
\multiput(60,13)(10,0){3}{\line(0,1){4}}
\multiput(60,23)(10,0){3}{\line(0,1){4}}
\multiput(70,33)(10,0){2}{\line(0,1){4}}
\end{picture}
}
\put(0,0){\begin{picture}(100,50)(-5,0)
\put(-5,45){{(c)}}
\multiput(5,5)(2,0){15}{\line(1,0){1}}
\multiput(5,5)(0,2){20}{\line(0,1){1}}
\multiput(5,45)(2,0){15}{\line(1,0){1}}
\multiput(35,5)(0,2){20}{\line(0,1){1}}
\put(10,10){\makebox(0,0){1}}
\put(20,10){\makebox(0,0){2}}
\put(30,10){\makebox(0,0){3}}
\put(10,20){\makebox(0,0){4}}
\put(20,20){\makebox(0,0){5}}
\put(30,20){\makebox(0,0){6}}
\put(10,30){\makebox(0,0){7}}
\put(20,30){\makebox(0,0){8}}
\put(30,30){\makebox(0,0){9}}
\put(10,40){\makebox(0,0){10}}
\put(20,40){\makebox(0,0){11}}
\put(30,40){\makebox(0,0){12}}
\multiput(3,10)(10,0){4}{\line(1,0){4}}
\multiput(3,20)(10,0){4}{\line(1,0){4}}
\multiput(3,30)(10,0){4}{\line(1,0){4}}
\multiput(3,40)(10,0){4}{\line(1,0){4}}

\multiput(10,3)(10,0){3}{\line(0,1){4}}
\multiput(10,13)(10,0){3}{\line(0,1){4}}
\multiput(10,23)(10,0){3}{\line(0,1){4}}
\multiput(10,33)(10,0){3}{\line(0,1){4}}
\multiput(10,43)(10,0){3}{\line(0,1){4}}
\put(45,25){\makebox(0,0){+}}
\multiput(55,5)(2,0){15}{\line(1,0){1}}
\multiput(55,5)(0,2){20}{\line(0,1){1}}
\multiput(55,45)(2,0){15}{\line(1,0){1}}
\multiput(85,5)(0,2){20}{\line(0,1){1}}
\put(60,10){\makebox(0,0){13}}
\put(70,10){\makebox(0,0){14}}
\put(70,20){\makebox(0,0){15}}
\put(60,20){\makebox(0,0){16}}
\put(80,10){\makebox(0,0){17}}
\put(80,20){\makebox(0,0){18}}
\put(80,30){\makebox(0,0){19}}
\put(70,30){\makebox(0,0){20}}
\put(60,30){\makebox(0,0){21}}
\put(80,40){\makebox(0,0){22}}
\put(70,40){\makebox(0,0){23}}
\multiput(53,10)(10,0){4}{\line(1,0){4}}
\multiput(53,20)(10,0){4}{\line(1,0){4}}
\multiput(53,30)(10,0){4}{\line(1,0){4}}
\multiput(63,40)(10,0){3}{\line(1,0){4}}

\multiput(60,3)(10,0){3}{\line(0,1){4}}
\multiput(60,13)(10,0){3}{\line(0,1){4}}
\multiput(60,23)(10,0){3}{\line(0,1){4}}
\multiput(60,33)(10,0){3}{\line(0,1){4}}
\multiput(70,43)(10,0){2}{\line(0,1){4}}
\end{picture}
}
\end{picture}
\caption{
Examples of the cluster consisting of two polymers ( $k=1$) 
that contributes to the lowest-order term of the expansion series 
for (a) $W(l_x=3,l_y=4;q=18)$, (b) $W(l_x=3,l_y=4;q=21)$ 
and (c)$W(l_x=3,l_y=4;q=24)$. 
}
\end{figure}
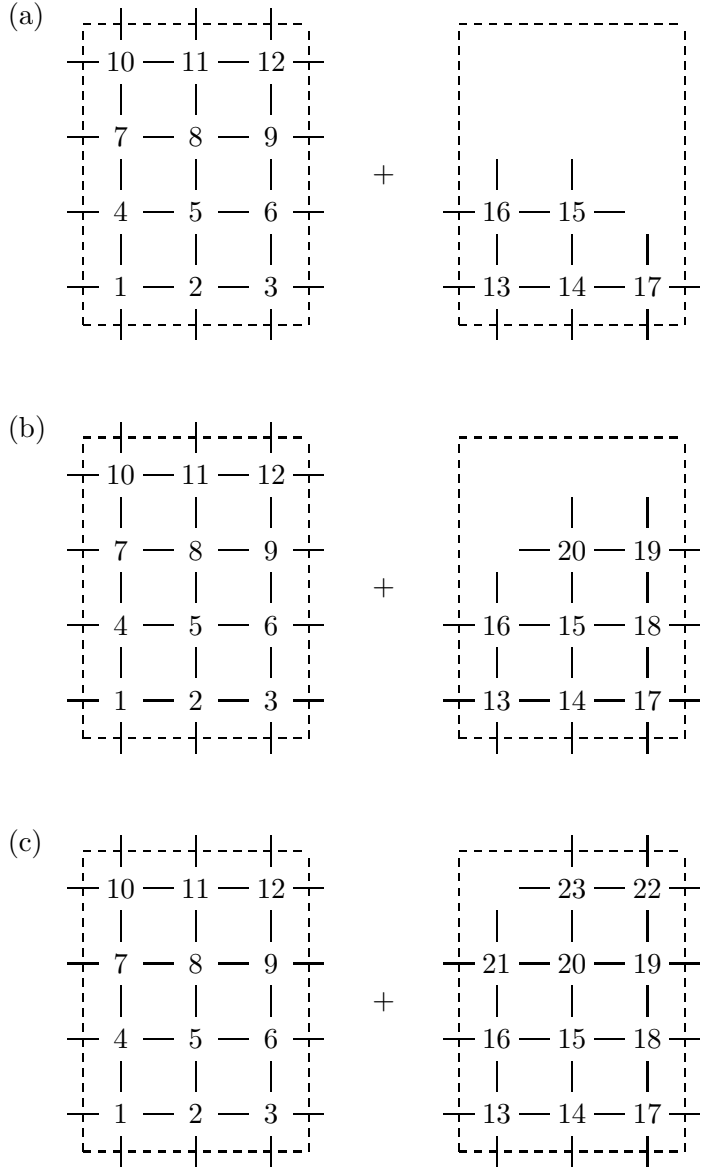
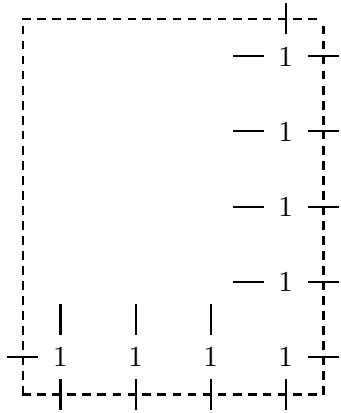
\begin{figure}
\setlength{\unitlength}{1mm}
\begin{picture}(110,60)
\put(30,0){\begin{picture}(50,60)(-5,0)
\multiput(5,5)(2,0){20}{\line(1,0){1}}
\multiput(5,5)(0,2){25}{\line(0,1){1}}
\multiput(5,55)(2,0){20}{\line(1,0){1}}
\multiput(45,5)(0,2){25}{\line(0,1){1}}
\multiput(10,10)(10,0){4}{\makebox(0,0){1}}
\multiput(40,20)(0,10){4}{\makebox(0,0){1}}
\put(3,10){\line(1,0){4}}
\multiput(10,3)(10,0){4}{\line(0,1){4}}
\multiput(10,13)(10,0){3}{\line(0,1){4}}
\multiput(33,20)(0,10){4}{\line(1,0){4}}
\multiput(43,10)(0,10){5}{\line(1,0){4}}
\multiput(40,53)(0,10){1}{\line(0,1){4}}
\end{picture}
}
\end{picture}
\caption{
Examples of the cluster consisting of a single polymer 
that contributes to the lowest-order term of the expansion series
for $\tilde{W}(l_x=4,l_y=5;\xi_Q)$. 
}
\end{figure}

\begin{figure}[t]
  \epsfxsize=15cm
  \centerline{\epsfbox{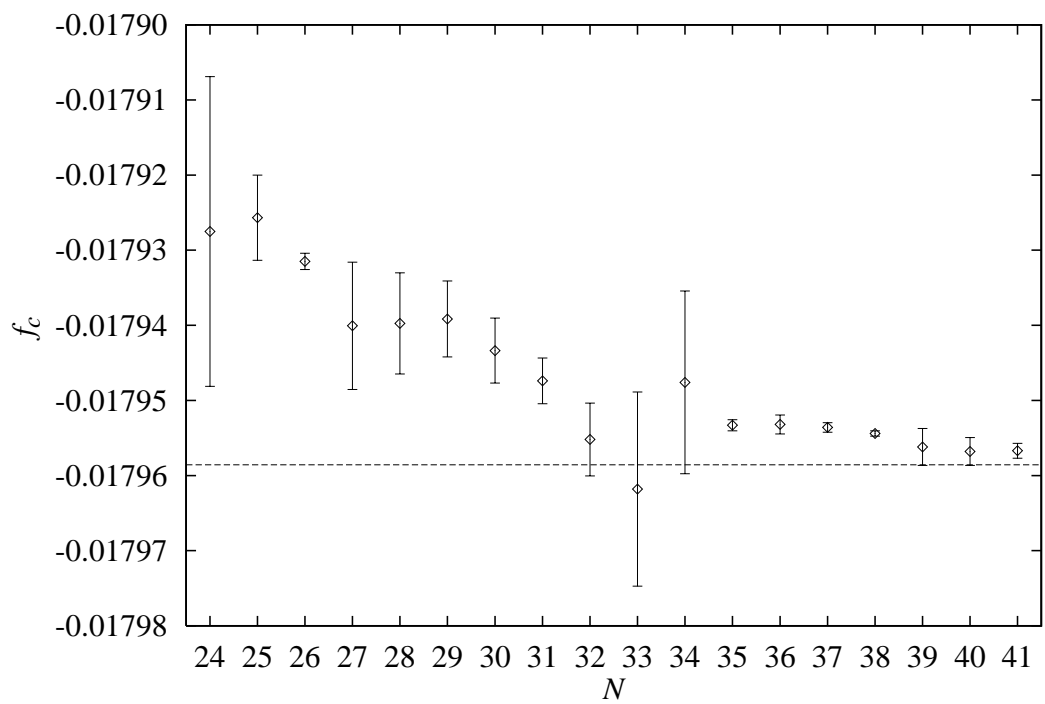}}
\caption{The estimates of the free energy density $f_c$ for $Q=20$ versus
the number of terms of the series. The solid line shows the exact
value.}
\end{figure}
\clearpage
\begin{figure}[t]
  \epsfxsize=15cm
  \centerline{\epsfbox{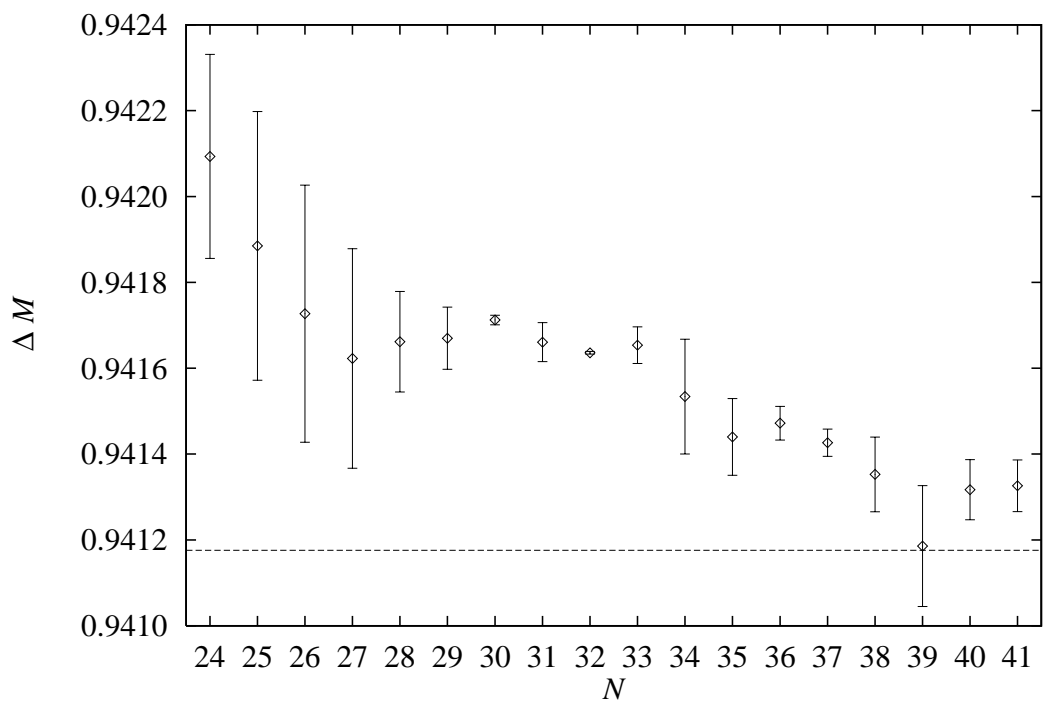}}
\caption{The estimates of the magnetization gap $\Delta M$ for $Q=20$ versus
the number of terms of the series. The solid line shows the exact
value.}
\end{figure}
\clearpage
\begin{figure}[t]
  \epsfxsize=15cm
  \centerline{\epsfbox{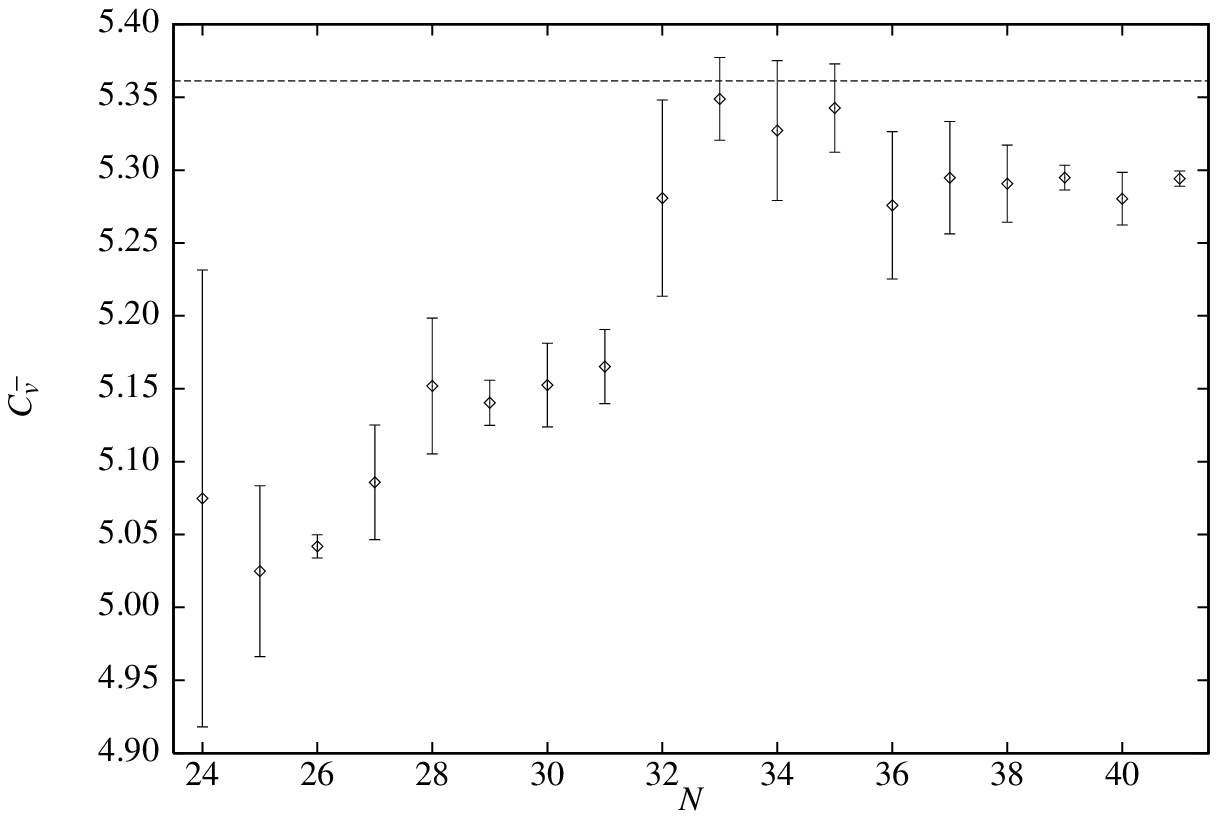}}
\caption{
The estimates of the specific heat $C_v^-$ for $Q=20$ versus
the number of terms of the series. The solid line shows the estimates
obtained from the large $Q$ expansion series by using the regularized
logarithmic Pad\'e approximants.
}
\end{figure}
\clearpage
\begin{figure}[t]
  \epsfxsize=15cm
  \centerline{\epsfbox{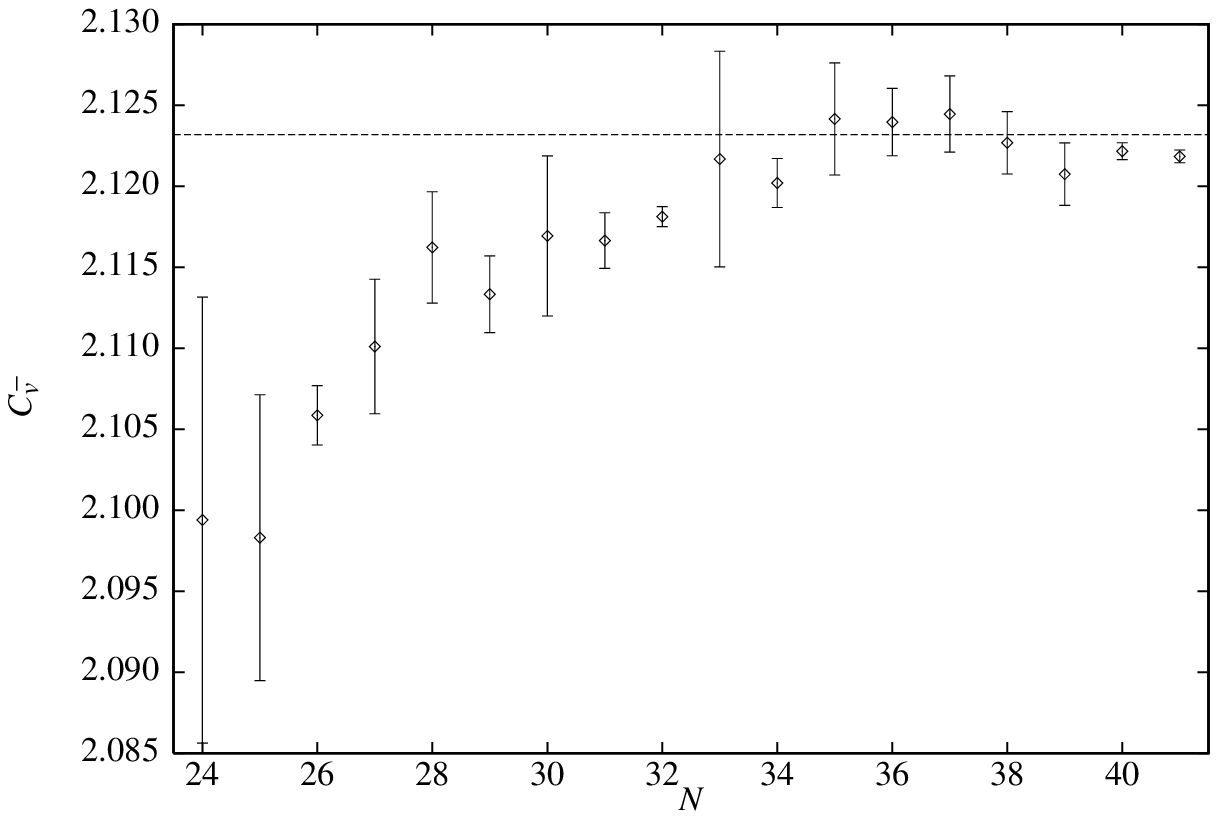}}
\caption{
The estimates of the specific heat $C_v^-$ for $Q=50$ versus
the number of terms of the series. The solid line shows the estimates
obtained from the large $Q$ expansion series by using the regularized
logarithmic Pad\'e approximants.
}
\end{figure}
\clearpage


\begin{thebibliography}{9}
\bibitem{Enting1977}
  T. de Neef and I. G. Enting, 
     J. Phys. A10 (1977) 801; \\
  I. G. Enting,
     J. Phys. A11 (1978) 563;
     Aust. J. Phys. 31 (1978) 515.
\bibitem{Creutz}
  M. Creutz, 
     Phys. Rev. B43 (1991) 10659.
\bibitem{Arisue1984}
    H. Arisue and T. Fujiwara, 
      Prog. Theor. Phys. 72 (1984) 1176; \\
      Preprint RIFP-588 (1985 unpublished); \\
    H. Arisue, 
      Nucl. Phys. B (Proc. Suppl.) 34 (1994) 240.
\bibitem{Enting1980}
  I. G. Enting,
     J. Phys. A13 (1980) 3713; \\
  G. Bhanot,
     J. Stat. Phys. 60 (1990) 55.
\bibitem{Arisue1995b}
  H. Arisue, 
     Nulc. Phys. B446[FS] (1995) 373.
\bibitem{Hasenbusch1994}
  M. Hasenbusch and K. Pinn, 
     Physica A203 (1994) 189.
\bibitem{Potts} 
  R. B. Potts, Proc. Camb. Phil. Soc. 48 (1952) 106.
\bibitem{Baxter} 
  R. J. Baxter,
     J. Phys. C6 (1973) L445;
     J. Phys. A15 (1982) 3329.
\bibitem{Bhanot1993} 
  G. Bhanot, M. Creutz, U. Gl\"assner, I. Horvath , J. Lacki,
  K. Schilling  and J. Weckel, 
     Phys. Rev. B48 (1993) 6183.
\bibitem{Address} 
     E-mail address: arisue@ipc.osaka-pct.ac.jp and\\
     URL of the web-site: http://www2.yukawa.kyoto-u.ac.jp/\~{}arisue/
\bibitem{Briggs}
  K. M. Briggs, I. G. Enting and A. J. Guttmann, 
     J. Phys. A27 (1994) 1503.       
\bibitem{Kihara}
  T. Kihara, Y. Midzuno and T. Shizume,
     J. Phys. Soc. Jap. 9 (1954) 681.
\bibitem{Straley}
  J. P. Straley and M. E. Fisher,
     J. Phys. A6 (1973) 1310.
\bibitem{Guttmann1993}
  A. J. Guttmann and I. G. Enting, 
     J. Phys. A26 (1993) 807.
\bibitem{Domb1974}
  C. Domb,
     Phase Transitions and Critical Phenomena vol.3 
     ed. C. Domb and M. S. Green (Academic, New York 1974). 
\bibitem{Muenster1981}
  G. M\"unster, 
     Nucl. Phys. B180[FS2] (1981) 23.
\bibitem{IDA}
  A. J. Guttmann and G. S. Joyce,
     J. Phys. A5 (1972) L81;
     Pad\'e approximants and their applications,
        eds. P. Graves-Morris (Academic Press, New York, 1973); \\
  G. L. Hunter and G. A. Baker,
     Phys. Rev, B19 (1979) 3808; \\
  M. E. Fisher and H. Au-Yang,
     J. Phys. A12 (1979) 1677.
\bibitem{Bhattacharya}
  T. Bhattacharya, R. Lacaze and A. Morel,
     Preprint SPhT-96/001 (1996).
\bibitem{Billoire92}
  A. Billoire, R. Lacaze and A. Morel,
     Nucl. Phys. B370 (1992) 773.
\bibitem{Billoire93}
  A. Billoire, T. Neuhaus and B. Berg,
     Nucl. Phys. B396 (1993) 779.
\bibitem{Janke92}
  W. Janke, B. A. Berg and M. Katoot,
     Nucl.Phys. B382 (1992) 649.
\bibitem{Rummukainen}
  K. Rummukainen,
     Nucl. Phys. B390 (1993) 621.
\bibitem{Janke95}
  W. Janke and S. Kappler,
     Europhys. Lett. 31 (1995) 345.
\end{thebibliography}
\end{document}